\documentstyle[11pt]{article}
\begin{document}
\pagestyle{plain}
\setcounter{page}{1}
\begin{center}
{\large\bf Variable Speed of Light Theories}
\vskip 0.3 true in
{\large J. W. Moffat}
\vskip  0.3 true in
Department of Physics, University of Toronto, Toronto, Ontario, Canada
\vskip 0.2 true in
and
\vskip 0.2 true in
Perimeter Institute for Theoretical Physics, Waterloo, Ontario, Canada
\vskip 0.2 true in
\begin{abstract}
Two variable speed of light models and their physical consequences are
investigated.
\end{abstract}
\end{center}
\vskip 0.2 true in
Talk given at the workshop on varying fundamental constants, JENAM 2002,
Porto, Portugal, September 2-7, 2002. To be published in the proceedings by
Kluwer publications.
\vskip 0.2 true in

{\bf 1. Varying Speed of Light Cosmology: An Alternative to Inflation}

\vskip 0.2 true in
It is ten years ago that an alternative solution to the
initial value problems of cosmology based on a variable speed of light
(VSL) was published~\cite{Moffat}. The model was based on the idea that in
the very early universe at a time $t\sim t_P\sim 10^{-43}$ sec., where
$t_P$ denotes the Planck time, the local Lorentz invariance of the ground
state of the universe was spontaneously broken by means of a non-zero vev
of a vector field, $\langle\phi^a\rangle_0 \not=0$, where $a$ labels the
flat tangent space coordinates of four-dimensional spacetime. At a
temperature $T < T_c$, the local Lorentz symmetry of the vacuum was
restored corresponding to a ``non-restoration'' of the symmetry group
$SO(3,1)$ as the temperature $T$ increases\footnote{Models of such a
non-restoration of symmetry in which a larger symmetry group breaks down
to a smaller one as the temperature increases and passes through a
critical temperature $T_c$ exist in the literature, see~\cite{Moffat2} for
references.}

Does there exist an
alternative to inflation, which can successfully allow a quantum field
theory calculation of a scale invariant primordial spectrum? In spite of
the successes of inflation theory, it is important to seek alternatives to
it to see whether a different scenario could overcome some of the
shortcomings of inflation, such as the problem of vacuum energy, the
fine-tuning of the coupling constant to give the correct density profile in
the present universe, and the unnaturally flat potentials needed to solve
the initial value problems.

In the following, we shall consider anew the
variable speed of light cosmology associated with a spontaneous symmetry
breaking of Lorentz invariance, and a phase transition in the speed of light
in the very early universe. A vierbein ${e_\mu}^a$ is used to convert
$\phi^a$ in flat tangent space into a 4-vector in coordinate space:
$\phi^\mu={e_a}^\mu\phi^a$. We introduce a variable speed of light with
$c(x)={\bar c}\chi(x)$, where ${\bar c}$ is a constant with
dimensions of velocity and $\chi(x)$ is a scalar field. The total action of
the theory is $S=S_G+S_M+S_\phi+S_\chi$, where
\begin{equation}
S_G=-\frac{c^4}{16\pi G}\int d^4xe(R+2\Lambda),
\end{equation}
$S_M$ is the matter action and
\begin{equation}
S_\phi=\int
d^4x\sqrt{-g}\biggl[\frac{1}{2}D_\mu\phi_aD^\mu\phi^a-V(\phi)\biggr].
\end{equation}
Moreover, the action $S_\chi$ is
\begin{equation}
\label{chiaction}
S_\chi=\int d^4x\sqrt{-g}\biggl[\frac{1}{2}D_\mu\chi
D^\mu\chi-V(\chi)-V(\chi\phi)\biggr],
\end{equation}
and $D_\mu$ is the covariant derivative
operator: $D_\mu=\partial_\mu\delta^a_b+(\Omega_\mu)^a_b$ where $(\Omega_\mu)^a_b$ is the
spin, gauge connection.\footnote{We demand that $\phi^a$ ($\phi^\mu$) be
a timelike vector which ensures that the kinetic energy term
$D_\mu\phi_aD^\mu\phi^a >0$ for all events in the past and future light
cones of the flat tangent space, which avoids the occurrence of negative
energy modes in the Hamiltonian. We could add a Lagrange multiplier term
to the action to guarantee the timelike nature of the vector $\phi^a$.
An alternative kinetic energy term would be $(\nabla_\mu\phi^\mu)^2$, which
has no negative energy states (private communication: M. A. Clayton).}

If the ``Mexican hat'' potential $V$ has a minimum at
$\phi_a=v_a$, then the spontaneously broken solution is given by
$v_a^2=\mu/4\lambda$. We can choose $\phi_a$ to be
$\phi_a=\delta_{a0}v=\delta_{a0}(\mu^2/4\lambda)^{1/2}$.
All the other solutions of $\phi_a$ are related to this one by a Lorentz
transformation. Then, the homogeneous Lorentz group $SO(3,1)$ is broken
down to the spatial rotation group $O(3)$. The three rotation generators
$J_i\,(i=1,2,3)$ leave the vacuum invariant, $J_iv_i=0$, while the three
Lorentz-boost generators $K_i$ break the vacuum symmetry, $K_iv_i\not= 0$.
In the spontaneously broken Lorentz symmetry phase, we can now
have the speed of light $c$ undergo a phase transition, since we are no
longer required to satisfy Einstein's second postulate of special
relativity, namely, that the speed of light is a constant with respect to
all local inertial frame observers.

Before the phase
transition at a time $t\sim t_c$, the radiation density and the entropy of
the universe, which are proportional to $c^{-3}$,  are reduced by many
orders of magnitude for $c\gg c_m$ (where $c_m=299792458\,m\,s^{-1}$ is the
currently  measured speed of light), allowing for a semiclassical quantum
field theory calculation of a scale invariant fluctuation spectrum. After
the phase transition has occurred, the radiation density and the entropy
of the universe increase hugely as $c\rightarrow c_m$, and the increase in
entropy follows the arrow of time determined by the spontaneously broken
direction of the vev $\langle\phi^a\rangle_0$. This solves the enigma of
the arrow of time and the second law of thermodynamics.

The phase transition in the speed of light with ${\dot c}/c_m <
0$ and $\log_{10}(c/c_m)\geq 30$ solves the horizon and flatness problems
with positive radiation pressure and density.

We shall consider a simple model of a free, minimally coupled
scalar field $\psi$, which we identify with our physical field $\psi$ in
the ``unitary gauge' after the three Goldstone modes have been removed in
our model of spontaneous symmetry breaking of Lorentz invariance. We choose
for simplicity flat spacetime with $k=0$.
The scalar field $\psi$ is pictured as a
plane wave mode with coordinate wave vector ${\vec k}$:
$\psi({\vec x},t)=\psi_k(t)\exp(i{\vec k}\cdot{\vec x})$,
which satisfies
\begin{equation}
\label{Harmonic}
{\ddot\psi}_k+3H{\dot\psi}_k+\frac{c^2k^2}{R^2}\psi_k=0,
\end{equation}
where have defined
$\psi_k=\frac{1}{(2\pi)^{3/2}}\int d^3x\psi({\vec x})\exp(-i{\vec
k}\cdot{\vec x})$.

We consider that the quantum fluctuation modes are created
in the spontaneously broken ground state and that their proper wavelength
$\lambda_p=R/k$ is tiny compared to the Hubble radius,
$R_H=c/H$~\cite{Moffat2}. Moreover, a possible solution of the equations of
motion for $\chi$ obtained from the action (\ref{chiaction}) is
$c(t)=a/t^b+c_0\theta(t_c-t)+c_m\theta(t-t_c)$. Our equation of motion for
$\psi_k$ is of the same form as the harmonic oscillator equation with a
unit mass, a variable spring constant $c^2k^2/R^2$, and a variable
friction damping coefficient $3H$. For ${\dot\psi}_k\sim H\psi_k$, a
radiation dominated universe ($p=\frac{1}{3}\rho$ and $H\sim 1/t$) and
$t\rightarrow 0$ the proper wavelength, $\lambda_p$, of a mode is much
smaller than the Hubble radius $R_H$, and the mode oscillates like an
ordinary harmonic oscillator with small damping. However, when $t$
increases $\lambda_p$ will eventually become greater than $R_H$, and the
mode enters an overdamped phase with ${\dot\psi}_k\sim 0$ and the mode
``freezes''. The ground state of the oscillator at some fixed time $t$ has
the form of a Gaussian wave function with a spread given by
$(\Delta\psi_k)^2=1/(2R^2ck)$.

For proper wavelengths much smaller than the Hubble radius, the
universe evolves adiabatically, whereas for wavelengths larger than the
Hubble radius, the overdamped modes cease to oscillate and $\Delta\psi_k$
will become constant. After the comoving wavelengths pass through the
horizon they freeze and the spectrum spread is given by
$(\Delta\psi_k)_h^2=1/(2R_h^2c_hk)$,
where $c_h$ and $R_h$ are the values of $c$ and $R$ at the time the
modes cross the Hubble radius, $R_H$, i.e. when
$R_h/k=c_h/H_h$.

The fluctuation modes at later times have the spectrum
$(\Delta\psi_k)_h^2\sim H_h^2/c_h^3k^3$.
This constitutes the prediction of a scale invariant spectrum with
$k^3\vert\delta_k\vert\sim {\rm constant}$,
where $\delta_k$ is the fractional energy density fluctuation in momentum
space.
We observe that at later times:
$\biggl(R/R_h\biggr)^2\biggl(c_0/c_h\biggr)(\Delta\psi_k)^2
=(\Delta\psi_k)_h^2.$
We see that for $c_h=c_m$, the spread $(\Delta\psi_k)^2$ is magnified by
the huge factor $(c/c_m)\sim 10^{30}$, so that the late time quantum
fluctuations have macroscopically relevant cosmological interest. In
inflation theory, it is the factor $(R/R_h)^2$ that is exponentially
enhanced and also produces macroscopically large fluctuation effects.

We have succeeded in obtaining a viable VSL model, which predicts a scale
invariant, Gaussian and adiabatic spectrum that agrees with the prediction
of inflation and the observational data. The semi-classical approximation
used in our quantum field theory calculation is justified, for the
radiation energy density and the entropy are severely diluted due to the
large value of $c$ before the phase transition.
A fundamental difference
between VSL cosmology and inflation is that we can choose the cosmological
constant $\Lambda$ to be small or zero from the beginning of the universe.
\vskip 0.2 true in

{\bf 2. Bimetric Gravity Theory, Dimming of Supernovae and Dark
Energy}
\vskip 0.2 true in
In the bimetric scalar-tensor gravitational theory there are two frames associated
with the two metrics ${\hat g}_{\mu\nu}$ and $g_{\mu\nu}$, which are linked
by the gradients of a scalar field: $\hat{g}_{\mu\nu}=g_{\mu\nu}
+B\partial_\mu\phi\partial_\nu\phi$~\cite{Clayton}.

The choice of a comoving frame for the
metric ${\hat g}_{\mu\nu}$ or $g_{\mu\nu}$ has fundamental physical consequences for
local observers in either metric spacetimes, while maintaining diffeomorphism
invariance. When the metric $g_{\mu\nu}$ is chosen to be associated with comoving
coordinates, then the speed of light varies in the frame with the metric ${\hat
g}_{\mu\nu}$. We call this the variable speed of light (VSL) metric frame.

If we choose ${\hat
g}_{\mu\nu}$ to be associated with comoving coordinates, then the speed
of light is constant but the speed of gravitational waves varies with
time (VSGW frame). The Friedmann equation in the VSGW frame is given by
\begin{equation}
\label{Kfriedmann}
H^2+\frac{c_m^2kK}{R^2}=\frac{8\pi G}{3}K^{3/2}\rho
+\frac{1}{3}c_m^2\Lambda K+\frac{1}{6}{\tilde\rho}_\phi,
\end{equation}
where we have defined:
${\tilde\rho}_\phi=\frac{1}{2}{\dot\phi}^2+c_m^2KV(\phi)$.
We see that when $v_g\equiv
c_mK^{1/2}=c_m[1-(B/c_m^2){\dot\phi}^2]^{1/2}\rightarrow 0$ ($v_g$ is the
speed of gravitational wave propagation) in the early universe, then
$H\rightarrow{\dot\phi}/\sqrt{2}$ and for ${\dot\phi}$ slowly varying,
this produces an inflationary solution in the initial universe, without a
``slow roll'' potential approximation as in standard inflationary
theories.

Observers in the VSL metric frame see the dimming of supernovae,
because of the increase of the luminosity distance versus red shift, due
to an increasing speed of light in the past universe, $\sim 10\%$ between $z=0$ and $z\sim 2-3$.
Moreover, in this frame the scalar field $\phi$ describes a dark energy
component in the Friedmann equation for the cosmic scale without
acceleration~\cite{Moffat3}. On the other hand, an observer in the VSGW
metric frame will observe the universe to be accelerating and the
supernovae will appear to be farther away.

Our bimetric gravity theory predicts that the gravitational
constant $G$ can vary in spacetime, {\it while the fine-structure constant,
$\alpha=e^2/\hbar c_m$, does not vary with time}, which allows for a $10-15\%$ increase
in the speed of light between $z=0$ and $z\sim 2-3$. The data of Webb et
al.~\cite{Webb}, showing a variation in $\alpha$ between $z=0.1$ and
$z=3.5$, is still to be confirmed by independent observational data.

\end{document}